# CFD SIMULATIONS OF TURBULENT FLOW IN THE HUMAN UPPER AIRWAYS

Elin AASGRAV[1*], Sverre G. JOHNSEN[2], Are J. SIMONSEN[2], Bernhard MÜLLER[1]

[1] NTNU, dept. Energy and Process Technology, 7491 Trondheim, NORWAY
[2] SINTEF Materials and Chemistry, 7465 Trondheim, NORWAY

* E-mail: eaasgrav@gmail.com

## ABSTRACT

In this paper, investigations are conducted using Reynolds-averaged Navier-Stokes (RANS) turbulence models to investigate the importance of turbulence modelling for nasal inspiration at a constant flow rate of 250 ml/s. Four different, standard turbulence models are tested in a model geometry based on pre-operative CT images of a selected obstructive sleep-apnea syndrome (OSAS) patient. The results show only minor differences between them. Furthermore, the turbulence models do not give significantly different results than a laminar flow model. Thus, the main conclusion is that effects of turbulence are insignificant in CFD modelling of the airflow in the pre-operative model of the upper airways of the chosen patient.

## NOMENCLATURE

*Greek Symbols*
$\delta_{ij}$    Kronecker delta, [-].
$\rho$    Mass density, [kg/m$^3$].
$\nu$    Kinematic viscosity, [m$^2$/s].
$\nu_T$    Turbulence eddy viscosity, [m$^2$/s].

*Latin Symbols*
$k$    Turbulence kinetic energy, [m$^2$/s$^2$].
$p$    Pressure, [Pa].
$U$    Velocity vector, [m/s].
$U_i$    Mean velocity component in the $i$ direction, [m/s].
$x, y, z$    Cartesian coordinates, [m].

*Sub/superscripts*
$i, j, k$    Spatial coordinate indexes.
$w$    Wall.

## INTRODUCTION

Snoring is caused by the soft parts of the upper airways collapsing and preventing the air from flowing freely. In some cases, snoring is so severe that medical attention is required. The most severe form, called obstructive sleep apnea syndrome (OSAS) involves complete blocking of the airway during sleep because of the collapse of e.g. relaxed muscles and soft tissue due to e.g. Venturi effect and gravity, in particular when the patient is lying in the supine position. It affects 2-4 % of the population. A variety of treatment options exists, but currently there are no available methods for predicting the outcome of the treatment. In order to gain insight into the biomechanical mechanisms of OSAS, computational fluid dynamics (CFD) simulations of flow in the human upper airways have been performed.

In short, the conclusions from previous studies indicate that the turbulence model that compares best with experimental data varies from case to case. Mihaescua et al. (2008) conclude that the Large Eddy Simulation (LES) modelling approach is a better option compared to the standard Reynolds-Averaged Navier-Stokes (RANS) models k-ε and k-ω, with k-ω being slightly better than k-ε. The RANS modelling approach is not able to capture flow separation effects, which are important for the understanding of the flow, as well as the LES approach. Riazuddin et al. (2011) conducted a study of inspiratory and expiratory flow in the nasal cavity using a k-ω SST turbulence model. The results were validated with experimental and numerical data from other studies, and they showed good correlation. The conclusion of the study was that the k-ω SST model gave accurate and reliable results for the flow involving adverse pressure gradients. Ma et al. (2009) used a realizable k-ε model when simulating flow and aerosol delivery in the human airways, and obtained good agreement with experimental data. Stapleton et al. (2000) used a standard k-ε model and concluded that CFD simulation do not compare very well with experimental data. They argued that the reason for this could be that particle deposition is very sensitive to pressure drop and recirculation, highlighting the need for accuracy in the reproduction of these flow characteristics to obtain good results. Longest et al. (2007) considered variations of the k-ω turbulence model. The standard k-ω model gave good agreement with experimental results, but a low-Reynolds number (LRN) k-ω model improved the results. They also emphasized the importance of accurate inlet conditions to obtain good results.

The studies all agree that CFD analysis of the human upper airways is a great tool for giving a realistic representation of flow related problems. Choosing a specific turbulence model can be challenging, because it depends, among other things, on the geometry and Reynolds number. The literature suggests that standard turbulence models are not always accurate enough, but improved models that take into consideration effects such as recirculation and separation, can provide results that agree well with empirical data. However, to our knowledge, no systematic studies have been published to compare and assess various turbulence models in the human upper airways (Quadrio et al., 2014).

The human upper airways consist of complex meatuses of highly varying cross-sections with hydraulic diameters ranging from milli- to centimeter-scale. Additionally, the sinusoidal nature of the intrathoracic pressure, due to the inhalation/expiration cycle, results in a wide range of flow velocities, hence Reynolds numbers. Most likely, the airflow is transitional, due to the relatively low maximum Reynolds number and the limited time to develop the turbulent boundary layers.

The current paper focuses on investigating the qualitative and quantitative differences between standard turbulence models applied to a patient-specific, rigid-wall geometry of the upper airways, investigated by Aasgrav (2016) based on CT images (Jordal, 2016). The study includes a sensitivity study with respect to grid size as well as turbulence boundary conditions. The present work is a part of the collaboration project "Modelling of obstructive sleep apnea by fluid-structure interaction in the upper airways" aiming to demonstrate the applicability of CFD as a clinical tool in OSAS diagnostics and treatment (OSAS, 2016). The project is a collaboration between NTNU, SINTEF and St. Olavs Hospital, the university hospital in Trondheim, and is funded by the Research Council of Norway.

**MODEL DESCRIPTION**

**Computational Geometry and Mesh of the Human Upper Airways**

The geometry retrieval is based on pre-operative CT scans of "Patient 12" (Moxness, 2014), provided by the Department of Radiology and Nuclear Medicine at St. Olavs Hospital, the university hospital in Trondheim. A detailed description of the process of retrieving the geometry can be found in the M.Sc. thesis by Jordal (2016). The resulting 3D geometry was modified to get an even distribution of outflow. The final pre-operational geometry used for further investigations is shown in Figure 1. The geometry has two inlets (left and right nostrils) and one outlet (trachea). The oral cavity was not considered in the model, and neither were the paranasal sinuses.

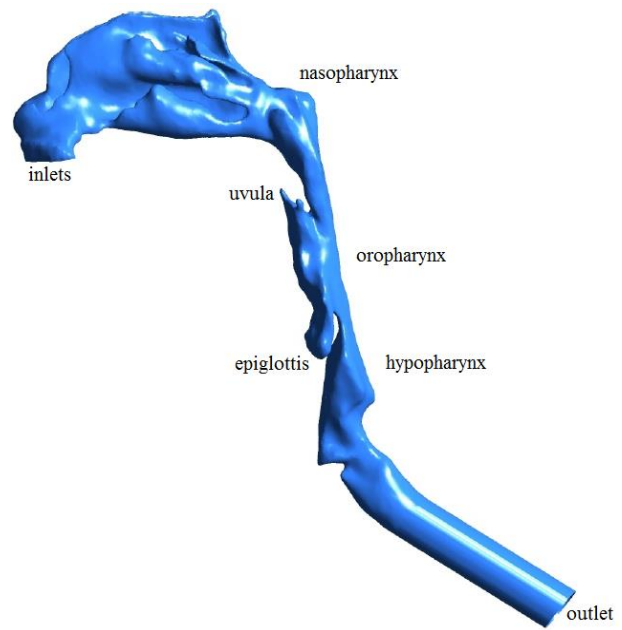

**Figure 1:** Final pre-operative model used in simulations, seen from the left (Jordal, 2016)

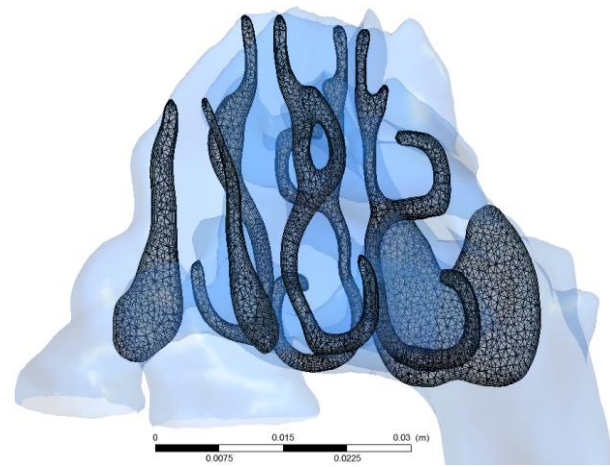

**Figure 2:** Base-case computational mesh in the nasal cavity, displayed on the cut-planes 1-4 (see Figure 4)

The meshing was done in ANSYS Meshing (Ansys, 2017), version 16.2. In order to get good results for the near-wall effects, an inflation layer consisting of five layers was utilized at the wall. The option "Size Function" in ANSYS Meshing was set to "Proximity and Curvature", where proximity captures the effects of tight gaps and thin sections, like for instance in the nasal cavity, and curvature captures sharp changes in flow direction, like we have in the nasopharynx. For the base-case, the size limitation was set to 1 mm. This resulted in a mesh with ca. 1.4 million grid cells. Details of the grid can be seen in figures 2 and 3. The grid sensitivity was investigated by comparing the base-case mesh to a refined mesh consisting of 6.8 million grid cells (size limitation of 0.8mm) and a coarser mesh consisting of 0.81 million grid cells (size limitation of 2.0mm), using the realizable k-ε turbulence model (see next section).

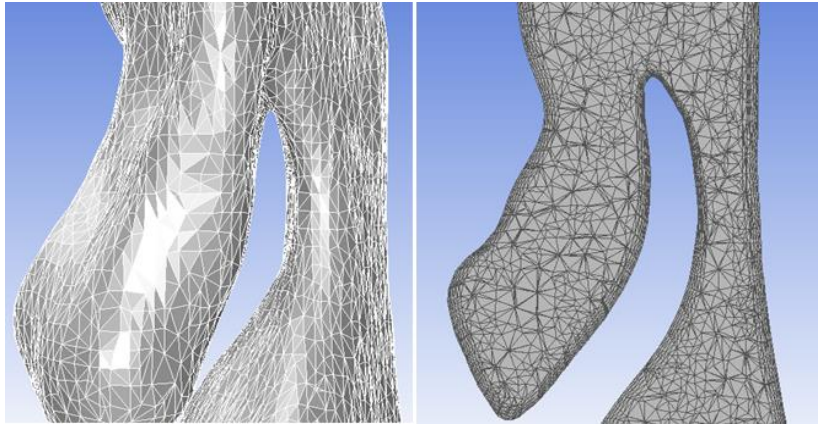

**Figure 3:** Details of the base-case mesh at the epiglottis; on the wall and in an arbitrary cut-plane

## Mathematical Models for Turbulent Flow in the Human Upper Airways

The Navier-Stokes equations describe fluid flow and thus are the foundation for the mathematical modelling of the airflow in the human upper airways. Due to the low Mach number (Ma<<0.3), the flow is considered incompressible, and the governing equations take the following form (Pope, 2000):

*Continuity equation*
$$\nabla \cdot \boldsymbol{U} = 0 \quad (1)$$

*Momentum equation*
$$\frac{D\boldsymbol{U}}{Dt} = -\frac{1}{\rho}\nabla p + \nu \nabla^2 \boldsymbol{U} \quad (2)$$

Here, $\boldsymbol{U}$ is the velocity vector, $p$ is the pressure, $\rho$ is the mass density and $\nu$ is the kinematic viscosity.

Even though we have a relatively low maximum Reynolds number of about 2000, we include effects of turbulence. Several solution approaches exist, with Reynolds-averaged Navier-Stokes (RANS) modelling being the most utilized one. Other popular methods are Large Eddy Simulation (LES) and Direct Numerical Simulation (DNS). DNS is solving the Navier-Stokes equations numerically for all significant spatial and temporal scales and does not involve any additional modelling of turbulence. LES involves explicit representation of the large-scale turbulent eddies containing anisotropic energy, while the smaller-scale, more isotropic turbulent motions are modelled. Although LES has a significantly lower computational cost than DNS, the RANS approach is far less computationally demanding. This makes RANS the desired approach in most practical cases. Here, we consider the RANS equations, where the Reynolds stress tensor is determined by the Boussinesq approximation.

*RANS equations*
$$\frac{\partial U_i}{\partial x_i} = 0 \quad (3)$$

$$\frac{\partial U_i}{\partial t} + U_j \frac{\partial U_i}{\partial x_j} = -\frac{\partial p}{\rho \partial x_i} + \nu \frac{\partial^2 U_i}{\partial x_j \partial x_j} - \frac{\partial \overline{u'_i u'_j}}{\partial x_j} \quad (4)$$

*Boussinesq approximation*
$$-\overline{u'_i u'_j} = 2\nu_T S_{ij} - \frac{2}{3} k \delta_{ij} \quad (5)$$

*Mean strain-rate tensor*
$$S_{ij} = \frac{1}{2}\left(\frac{\partial U_i}{\partial x_j} + \frac{\partial U_j}{\partial x_i}\right) \quad (6)$$

Here, $U_i$ and $U_j$ are the mean velocity components in the $i$ and $j$ directions, respectively ($i, j \in \{x, y, z\}$), $p$ is the mean pressure, $k$ is the turbulence kinetic energy, $\nu_T$ is the eddy viscosity to be defined by the RANS model, $\delta_{ij}$ is the Kronecker delta, and the Einstein summation convention is employed. The Reynolds stress models are generally divided into categories based on how many equations need to be solved, with the two-equation models being the most used and the most verified RANS types.

## Numerical Approximation

The governing equations were solved using the commercial CFD software ANSYS Fluent 16.2 (Ansys, 2017). In the following, simulation results from the upper airways geometry shown in the previous section are shown, for various standard RANS turbulence models as well as laminar flow. Coupled solver was employed for the pressure-velocity coupling. For pressure and momentum, second order upwind solvers were chosen, while for the turbulent kinetic energy and turbulent dissipation rate, a first order upwind solver was determined to be accurate enough. Standard material properties for air was employed (mass density of 1.225 kg/m$^3$ and viscosity of 1.7894·10$^{-5}$ Pa s).

Boundary conditions were:
  - Atmospheric total pressure at the inlets (nostrils)
  - Velocity outlet corresponding to an inspiratory volumetric flow rate of 250 ml/s
  - No-slip condition at the walls
  - Turbulence intensity of 5%
  - Turbulent viscosity ratio of 10

The sensitivity to turbulence boundary conditions at the inlets were investigated by testing the sensitivity to reducing the turbulence intensity at the inlets to 1% and increasing it to 10%.

**RESULTS**

The described setup was simulated with four different turbulence models, namely the standard k-ε and k-ω models, as well as realizable k-ε and k-ω SST. The four models were checked against laminar flow by comparing the area-averaged pressure at selected cross-sections throughout the geometry (see Figure 4). The results are shown in Figure 5. The models showed only minor differences in results upstream of the epiglottis. Some differences are observed downstream of the epiglottis, but it is believed that the effects so far down do not affect the flow further up where the airway collapses in OSAS.

It is to be expected that the difference between a laminar model and various turbulence models is minor, because the maximum Reynolds number in the flow is about 2000, indicating that the flow is mainly laminar. Because of the complex geometry inducing separated flow, the flow most likely has some turbulence features as well. The total pressure, $p_{tot} = p_{stat} + 0.5\rho U^2$, decreases throughout the geometry as it should, while the static pressure depicted in Figure 5 does not show this behavior for all the models because of the highly varying velocity.

Both of the k-ε models' residuals converged to an acceptable value, where the residuals for continuity, k, ε, and x-, y- and z-velocities started at about 1, and converged to values between $10^{-4}$ and $10^{-8}$, with a steady state solver. None of the k-ω models' residuals converged as desired with steady state. Thus, a transient simulation was needed to achieve residuals in the range of $10^{-4}$-$10^{-8}$. Despite the steady-state boundary conditions, the solution might be transient due to unsteady vortices in regions with separated flow. In this case, a converged steady-state solution would be unfeasible.

The realizable k-ε model was chosen for the grid and turbulence boundary condition sensitivity studies. First, a sensitivity study was conducted to investigate the sensitivity to turbulent intensity at the inlets, as described in the Model Description chapter. In Figure 6, it is seen that the turbulent kinetic energy differences that exist close to the inlets, due to the different turbulence intensity boundary conditions, decay as the air progresses through the nasal cavity, such that the effect of changing the inlet boundary condition is negligible when considering the flow entering the nasopharynx. Furthermore, we found that the velocity streamlines and velocity magnitude are largely unaffected by the turbulence intensity. Second, a grid sensitivity study as described in the previous chapter was performed utilizing a base-case grid (1.4M grid cells), a refined grid (6.8M grid cells) and a coarser grid (0.81M grid cells). Figure 7 shows a comparison of the area-averaged pressure at the selected cross-sections for the different grids. It is evident that the coarsest mesh differs from the base case and the finer mesh, leading to the conclusion that grid independency is achieved for the base case sizing and finer resolutions. The velocity streamlines in Figure 8 show that the three different meshes give some differences in the flow patterns. This is especially prominent right after the epiglottis and in the oropharynx behind the oral cavity. Here, the refined mesh portrays more swirl in the flow, indicating higher vorticity in these regions, a characteristic of the flow pattern that could be an important factor in the understanding of OSAS. The coarser mesh has less swirl than the two other meshes, indicating that the mesh is too coarse to capture the complexity and turbulence effects of the flow. The wall pressure was found not to show any difference between the three meshes. The turbulence kinetic energy plot was similar for the two finer meshes, while it differed greatly for the coarser mesh, giving the same conclusion that the coarser mesh does not have a large enough resolution to capture the turbulence effects.

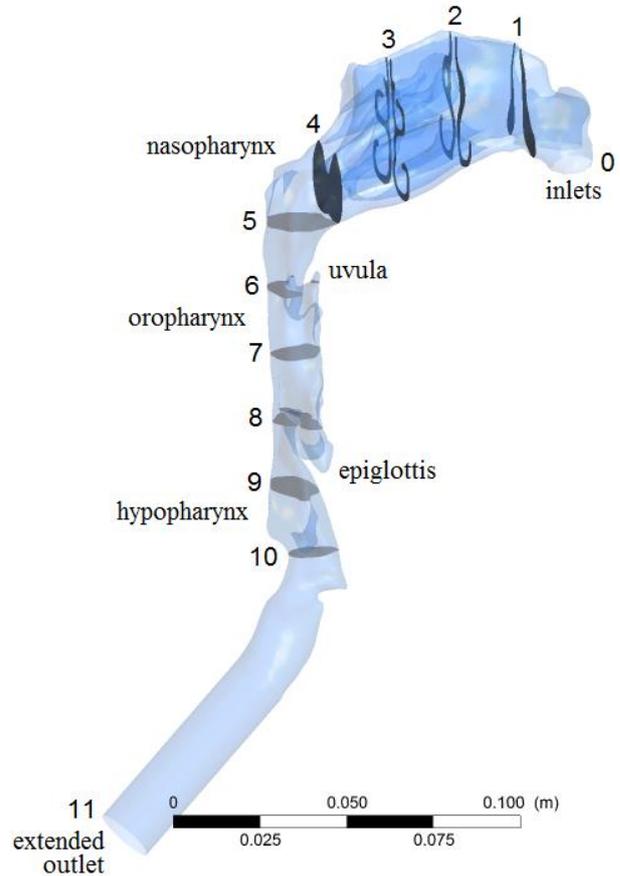

**Figure 4:** Location and numbering of cross-sections in the final pre-op geometry

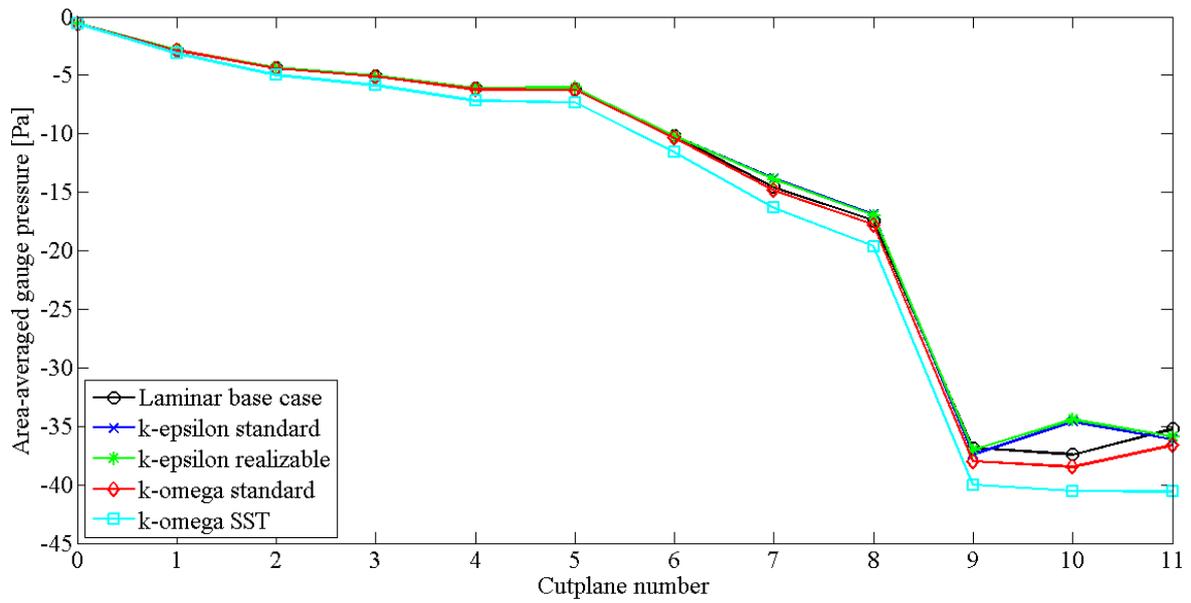

**Figure 5:** Comparison of area-averaged pressure for the laminar base-case and four different turbulence models

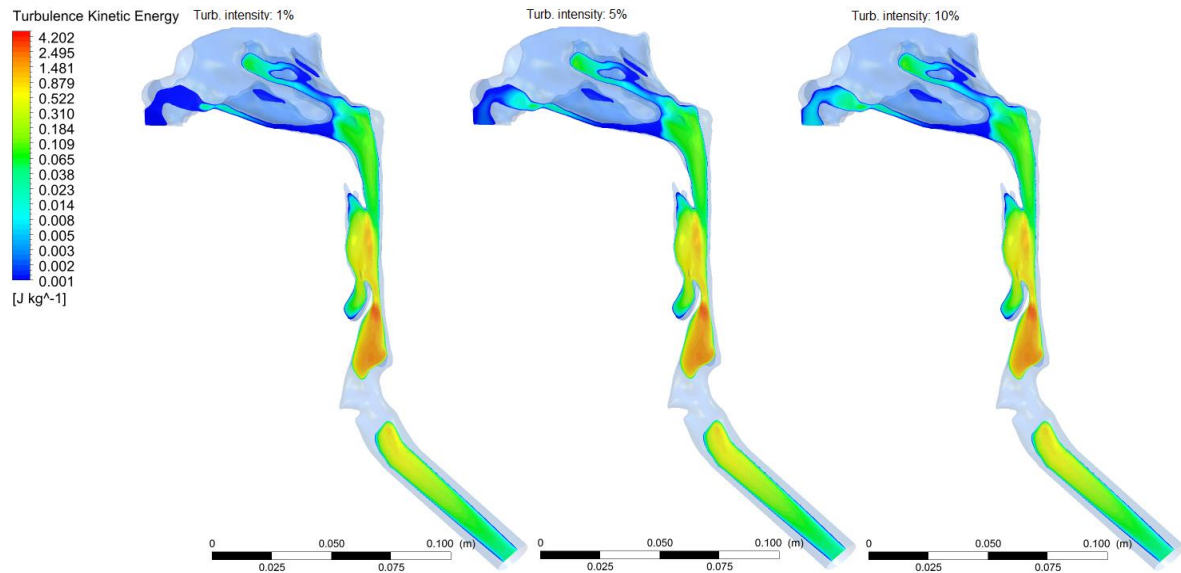

**Figure 6:** Turbulence kinetic energy for different turbulent intensities using the base case mesh, logarithmic scale

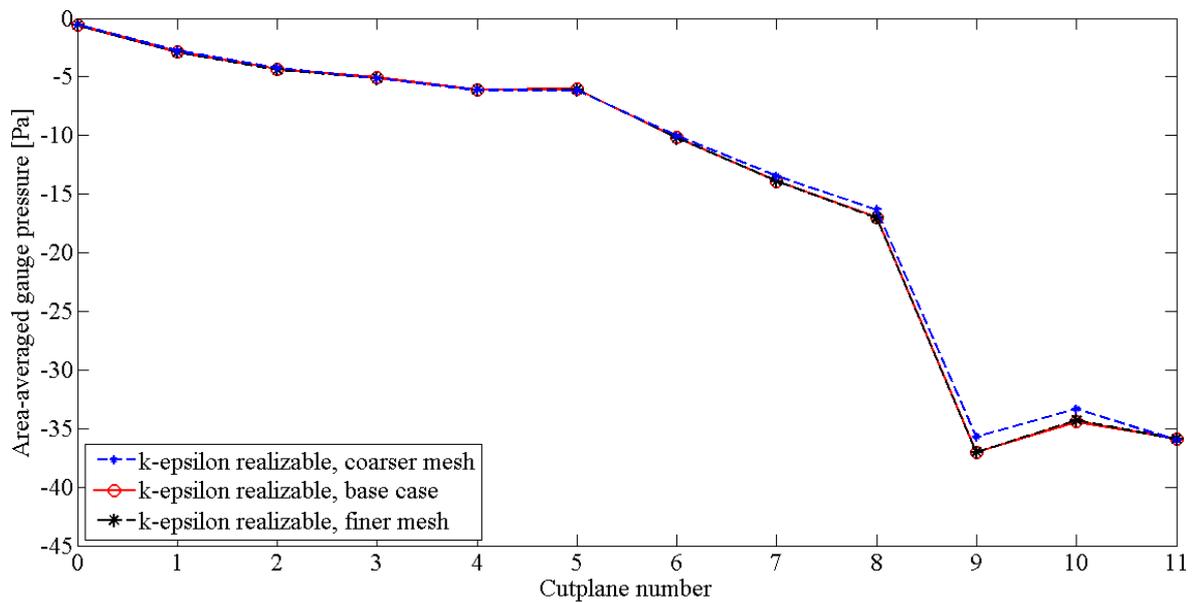

**Figure 7:** Comparison of area-averaged pressure for the turbulent base-case, and a finer and coarser mesh

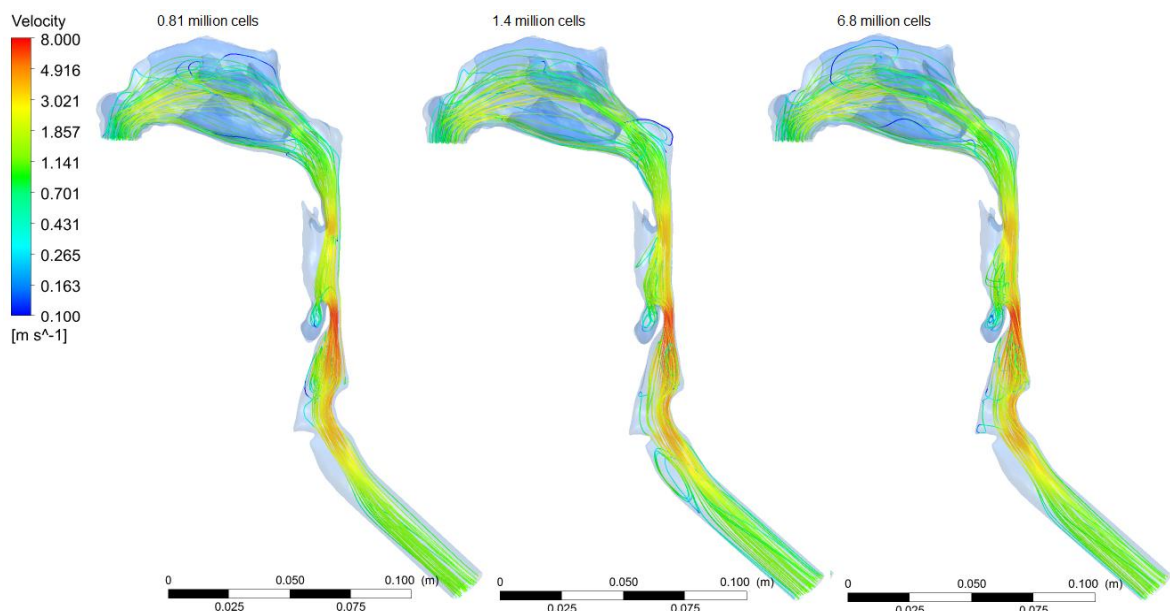

**Figure 8:** Velocity streamlines for different meshes, logarithmic scale

## CONCLUSIONS

CFD simulations of airflow in the human upper airways were performed to investigate and assess the importance of turbulence modelling. Four different standard RANS turbulence models were compared to a laminar flow model at a constant inspiratory volumetric flow rate of 250 ml/s in a model geometry based on pre-operative CT images of an OSAS patient. The area-averaged pressure at selected cross-sections upstream of the epiglottis were largely unaffected by the choice of laminar or turbulent flow models. Thus, the main conclusion of the study is that effects of turbulence are insignificant in CFD modelling of the airflow in the pre-operative model of the upper airways of the chosen patient. It remains to investigate other volumetric flow rates.

Employing the realizable k-ε model, the effect of varying turbulence inlet boundary conditions was investigated by varying the turbulent intensity at the inlets from 1% to 10%. No significant effect was observed downstream of the nasal cavities.

Finally, a grid sensitivity study was conducted to assess the grid independency of the computed results. The base-case mesh, based on a cell size limitation of 1 mm and consisting of 1.4 million cells, showed some discrepancy in the flow pattern in some regions, but produced almost exactly the same pressure loss results as a refined mesh consisting of 6.8 million cells (size limitation of 0.8 mm). A coarser mesh consisting of 0.81 million cells was not able to reproduce the results and thus did not have the required resolution to capture the turbulence effects.